\documentclass[
reprint,
showpacs,preprintnumbers,
 amsmath,amssymb,
 aps,
prd
]{revtex4-1}

\usepackage{dcolumn}
\usepackage{bm}
\usepackage[mathlines]{lineno}

\newcommand{\diff}{\widetilde{\rm Diff}^+(S^1)}
\newcommand{\diffc}{{\rm Diff}_c(S^1)}
\newcommand{\vectc}{{\rm Vect}_c(S^1)}
\newcommand{\diffext}{\widehat{\rm Diff}_c(S^1)}
\newcommand{\vectext}{\widehat{\rm Vect}_c(S^1)}

\numberwithin{equation}{section}

\begin{document}


\title{Inflationary Cosmology, Diffeomorphism Group of the Line and \\ Virasoro Coadjoint Orbits}

\author{James E. Lidsey}
\email{james.lidsey@aira.astro.ro}

\affiliation{
Astronomical Institute of the Romanian Academy, 
\\
Astronomical Observatory Cluj-Napoca, 
Str. Ciresilor 19, 400487 Cluj-Napoca, Romania}


\begin{abstract}
The cosmological field equations sourced by a self-interacting scalar field are dynamically equivalent to a closed system of equations obtained by applying the moment method to non-linear Schr\"odinger equations possessing an underlying non-relativistic conformal $SL(2,\mathbb{R})$ symmetry. We consider the one-dimensional, quintic Schr\"odinger equation relevant to strongly repulsive, dilute Bose gases. The action of the diffeomorphism group on the space of Schr\"odinger operators generates an harmonic trapping potential that can be identified with the kinetic energy of the cosmological scalar field. Inflationary cosmologies are represented by points on the orbit of the de Sitter solution, which is the quotient manifold ${\rm Diff} (\mathbb{R})/SL(2,\mathbb{R})$. Key roles are played by the Schwarzian derivative of the diffeomorphism and the Ermakov-Pinney equation. The underlying $SL(2,\mathbb{R})$ symmetry results in a first integral constraint which ensures energy-momentum conservation. When the analysis is restricted to the universal cover group of diffeomorphisms on the circle, the generation of a rolling scalar field can be understood in terms of the Virasoro coadjoint action. The corresponding symplectic two-form and Hamiltonian generator of the coadjoint orbit are determined by the scalar field kinetic energy.



\end{abstract}


\maketitle


\nopagebreak

\renewcommand{\thesection}{1}

\section{\label{introsection} Introduction}

A wealth of astrophysical observations, most notably from the cosmic microwave background \cite{Planck2015} and large-scale structure surveys \cite{DES2017}, strongly indicate that the universe underwent a phase of accelerated, inflationary expansion in its most distant past \cite{inflation} and, moreover, is currently entering a second accelerated phase at the present epoch \cite{darkenergy}. The simplest mechanism for driving such an expansion (at both early- and late-times in the universe's history) is through the potential energy associated with the self-interactions of a scalar field which is minimally coupled to Einstein gravity. (For reviews see, e.g., \cite{infreview,dereview}). Studying the dynamics of spatially isotropic cosmologies sourced by such a field is therefore of considerable importance. 

In the present work, we illustrate how inflationary dynamics can be described in terms of the group ${\rm Diff}^+(\mathbb{R})$ of orientation-preserving diffeomorphisms of the line corresponding to general temporal reparametrizations. Such a description is achieved by bringing together two apparently unrelated results: namely, an equivalence that exists between natural ${\rm Diff}^+(\mathbb{R})$ representations on the vector spaces of Schr\"odinger and Sturm-Liouville differential operators with time-dependent (harmonic trapping) potentials \cite{RU2006,U2010,UR2008}, and a (formal) correspondence between the dynamics of certain non-relativistic conformal field theories (NRCFTs) and the cosmological Einstein equations \cite{L2003}. The latter equivalence is established by identifying the integral moments of a non-linear Schr\"odinger wavefunction with appropriate cosmological parameters. The (cosmic) dynamics of the wavefunction moments is then governed by the Ermakov-Pinney equation \cite{E1880,M1930,P1950}. 

The unifying object that links together these apparently disparate themes is the Schwarzian derivative operator of the diffeomorphism \cite{OT2004} and, in particular, the identification of this operator with the kinetic energy of the cosmological scalar field. An essential role is also played by the non-relativistic conformal group $SL(2,\mathbb{R})$. The NRCFT we consider is the $d=1$ quintic, non-linear Schr\"odinger equation, which is invariant under the action of the Schr\"odinger group, $Sch(d)$, the maximum kinematical group of the free Schr\"odinger equation \cite{N1972,H1972}. This group is the semi-direct product $Sch(d)=G \triangleright SL(2,\mathbb{R} )$, where $G$ is the Galilean group with central extension and $SL(2,\mathbb{R})$ is the group of fractional linear transformations on the time coordinate. The underlying $SL(2,\mathbb{R})$ symmetry within the Schr\"odinger group ensures that the dynamics of the wavefunction integral moments form a closed, non-autonomous system of ordinary differential equations (ODEs) \cite{PTM2005}. The solution of this symmetric system can be interpreted in a cosmological context as the de Sitter (inflationary) solution.

The ${\rm Diff}^+(\mathbb{R})$-action explicitly breaks the conformal symmetry and generates an effective harmonic trapping potential in the Schr\"odinger equation, which is proportional to the Schwarzian derivative of the diffeomorphism \cite{G2001,RU2006}. The key point is that the space of Schr\"odinger operators is a ${\rm Diff}^+(\mathbb{R})$-module \cite{RU2006,U2010}, which implies that closure of the moment equations is preserved under ${\rm Diff}^+(\mathbb{R})$. The trapping potential can be identified in terms of a dynamical scalar field which evolves on a spatially isotropic spacetime. In this sense, a scalar field (inflationary) cosmology is generated by the action of ${\rm Diff}^+(\mathbb{R})$ and may be viewed as a point on the ${\rm Diff}^+(\mathbb{R})$-orbit of the pure de Sitter background. This orbit is the quotient manifold ${\rm Diff}^+(\mathbb{R})/SL(2,\mathbb{R})$. 

When the ${\rm Diff}^+(\mathbb{R})$-action is restricted to the subgroup corresponding to the universal cover group of orientation-preserving diffeomorphisms on the circle, the time-reparametrization can be understood in terms of the coadjoint representation of the Virasoro-Bott group \cite{LP1975,K1980,S1981,K1982,W1988}, which is the unique (up to isomorphism) central extension of ${\rm Diff}^+(S^1)$ \cite{GF1968,B1977}. 
A non-trivial central charge is essential for generating a rolling scalar field. The orbit of cosmologies in this case is a symplectic manifold and the associated non-degenerate, closed two-form, together with the corresponding Hamiltonian generator, can be calculated explicitly. The value of the two-form at each point on the orbit is determined by the kinetic energy of the scalar field and therefore quantifies the deviation away from the de Sitter point in the slow-roll inflationary scenario. 

Non-linear Schr\"odinger (NLS) equations play a key role in diverse branches of physics ranging from optics, plasma physics through to Bose-Einstein condensation \cite{F2015}. The quintic NLS equation is of particular relevance to one-dimensional ultra-cold, dilute Bose gases \cite{KNSX2000}. In the limit of strong, repulsive, interatomic interactions, such gases are in the so-called Tonks-Gerardeau regime, where the bosonic particles behave as a system of free fermions \cite{T1936,G1960}. The Tonks-Gerardeau regime can be (partially) described by the quintic NLS equation with an harmonic trapping potential, where the non-linearity arises as a consequence of many-body effects on the scattering cross-section of the particles \cite{KNSX2000,LBM2003,AG2004,CFK2008}. Tonks-Gerardeau gases have been realized experimentally \cite{Petal2004,KWW2004}. We keep this connection with ultra-cold atomic gases in mind throughout the course of this work. One-dimensional Bose gases trapped in periodic harmonic potentials were recently classified in terms of centrally extended Virasoro coadjoint orbits \cite{SK2017,SUK2018}. 

The cosmological implications of the Ermakov-Pinney  equation have been investigated by a number of authors \cite{R1998,RS1999,RER1999,RE1999,HL2002,WK2003,WKCHP2004,HKWCF2007,dAW2010,dA2010,L2013,CGGH2018}, and commentaries on its mathematical properties and numerous physical applications are provided in \cite{LA2008,ML2015,CGGH2018}. 
The possible role of the Schwarzian derivative in cosmology has been discussed recently. In \cite{G2014,C2015,DGH2016}, projective transformations of the time coordinate were employed to relate Newtonian cosmological backgrounds with a constant dark energy density to those where the density is time-dependent. Virasoro coadjoint orbits have been discussed in a quantum cosmological context in \cite{HR1994}. 

Finally, before proceeding, we should be clear about how the diffeomorphism group on the line should be interpreted in this work from a gravitational perspective. Diffeomorphisms in Einstein gravity are typically viewed as gauge redundancies, but this is not the case here. Rather, the natural representations of ${\rm Diff}^+(\mathbb{R})$ are equivalence groups on the spaces of differential Schr\"odinger and Sturm-Liouville operators within which the cosmological Friedmann equations can be embedded.  

The paper is organized as follows. In Section \ref{diffsection}, the relevant representations of ${\rm Diff}^+(\mathbb{R})$ are summarized and the important role played by tensor-densities is highlighted. We proceed in Section \ref{momentsection} to derive the wavefunction moment equations. The cosmological implications of the preceeding Sections are discussed in Section \ref{cosmosection}, specifically the map to the traceless Einstein field equations (Subsection \ref{frwsection}), the interpretation of inflationary cosmologies as points on the de Sitter orbit (Subsection \ref{orbitsection}), and the self-similar characteristics of the underlying wavefunctions (Subsection \ref{similarsection}). Some specific models based on exponential self-interaction potentials are also presented (Subsection \ref{examplesection}). The connection with Virasoro coadjoint orbits is discussed in Section \ref{virasorosection}. We conclude with a discussion in Section \ref{concsection}. A short Appendix is included for the benefit of readers unfamiliar with centrally extended Lie groups and coadjoint representations. 

Throughout this work, the vector space of smooth (continuously differentiable) 
functions on a manifold ${\cal{M}}$ is denoted by $C^{\infty}({\cal{M}})$. Depending on the context, derivatives of a function $y(x) \in C^{\infty}({\cal{M}})$ are denoted $dy/dx = \dot{y} (x) = y'(x) = \partial_x y$, etc. Unless stated otherwise, we denote the orientation-preserving group of diffeomorphisms of the one-dimensional manifold ${\cal{M}}$ by  ${\rm Diff}({\cal{M}})$, where ${\cal{M}}=\mathbb{R},S^1$. 
Physical units are chosen such that $c=\hbar =1$. 

\renewcommand{\thesection}{2}

\section{\label{diffsection} Representations of the Diffeomorphism Group}

\subsection{\label{schsection} On the Space of Schr\"odinger Operators}

It can be proved \cite{OG2012} that all cubic-quintic, non-linear Schr\"odinger (NLS) equations which possess a six-dimensional Lie point symmetry can always be transformed into the pure quintic equation:
\begin{equation}
\label{purequintic}
i\partial_t u + \frac{1}{2m}\partial_{rr} u -g_0 |u|^4 u =0
\end{equation}
where $g_0\in \mathbb{C}$ is a constant coupling and $m \in \mathbb{R}$ is a mass parameter. The $d=1$ Schr\"odinger group admits a six-dimensional Lie algebra, $sch_1$, and the Lie point symmetry of (\ref{purequintic}) is isomorphic to this algebra \cite{OG2012}. A representation for the $sl(2,\mathbb{R})$ subalgebra of $sch_1$ 
can be given in terms of first-order operators on $\mathbb{R}^2$ (spanned by coordinates $(t,r)$). It takes the form \cite{HU2003}
\begin{eqnarray}
\label{sch1rep}
L_{-1} & = & -\partial_t , \qquad L_0 =-t\partial_t -\frac{1}{2} r\partial_r - \mu  \nonumber \\
L_1 & = & - t^2 \partial_t -tr \partial_r -\frac{M}{2} r^2 -2\mu t 
\end{eqnarray}
where $M :=-im$ and $\mu \in \mathbb{R}$ is the scaling dimension of the wavefunction. (This should be set to $\mu=1/4$ when interpreting $u$ as a solution to
the $d=1$ Schr\"odinger equation (\ref{purequintic}).) 
 
This $sl(2,\mathbb{R})$ algebra may be embedded in an infinite-dimensional, centerless Virasoro-type algebra, ${\rm Vect}({\cal{M}}) :=\langle L_n \, |\, n \in \mathbb{Z} \rangle$, where the generators are represented by the first-order operators 
\begin{eqnarray}
\label{svrep}
L_n =-t^{n+1}\partial_t 
-\frac{n+1}{2} t^n r \partial_r \nonumber \\
-\mu (n+1)t^n 
-\frac{n(n+1)}{4}M t^{n-1}r^2 
\end{eqnarray}
These satisfy the non-vanishing commutation relations
\begin{equation}
\label{svbracket}
\left[ L_n ,L_{n'} \right]  =  (n-n' ) L_{n+n'} ,\qquad n,n'\in \mathbb{Z}
\end{equation}
and the $sl(2,\mathbb{R})$ subalgebra of $sch_1$ is spanned by $\langle L_n | n \in \{ 0,\pm 1 \} \rangle$. The algebra ${\rm Vect}({\cal{M}})$ is the Lie algebra of the group of diffeomorphisms ${\rm Diff}({\cal{M}})$ and is a subalgebra of the infinite-dimensional Schr\"odinger-Virasoro algebra \cite{H1993,HU2003,RU2006,U2010,UR2008,UR2012}.  

The infinitesimal transformations generated by (\ref{svrep}) may be formally exponentiated to yield the finite transformation laws for the coordinates $(t,r,u) \mapsto (\tau , s , \tilde{u} )$ \cite{HU2003}: 
\begin{eqnarray}
\label{virfinite}
t& = & \beta (\tau ) , \qquad r = ( \dot{\beta} (\tau ) )^{1/2} s
\\
\label{virwave}
u(t,r) &  = & ( \dot{\beta}(\tau ) )^{-\mu} \exp \left[ \frac{im}{4} \frac{\ddot{\beta}(\tau)}{\dot{\beta} (\tau )} s^2 
\right] \tilde{u}(\tau ,s)
\end{eqnarray}
where $\beta \in C^{\infty}({\cal{M}})$. The necessary condition $\dot{\beta}(\tau) >0$ restricts the transformation to the orientation-preserving subgroup of ${\rm Diff}({\cal{M}})$. Moreover, elements of ${\rm Diff} ({\cal{M}})$ are labeled by the map $f: t \mapsto f(t)$, but since this map is diffeomorphic, it is often more convenient to consider the inverse element $f^{-1} (\tau) := \beta (\tau)$. Eq. (\ref{virwave}) is a representation of the group ${\rm Diff}({\cal{M}})$ on a smooth function of two variables, where 
$\beta^{-1} \in {\rm Diff}({\cal{M}})$ induces the coordinate change (\ref{virfinite})
on $\mathbb{R}^2$. We denote this representation by $(\Pi_{\mu}(\beta )h)(t,r)$ acting on a function $h(t,r)\in C^{\infty}(\mathbb{R}^2)$.

When restricted to the action of $SL(2,\mathbb{R})$, the diffeomorphism is generated by (\ref{sch1rep}) and given explicitly by 
\begin{equation}
\label{sl2r}
\beta (\tau ) =\frac{a \tau +b}{c \tau +d} , \qquad ad-bc =1 
\end{equation}
This transformation leaves Eq. (\ref{purequintic}) invariant.
Clearly, the more general coordinate transformations (\ref{virfinite})-(\ref{virwave}) break this Lie point symmetry. Nonetheless, they constitute the equivalence group, $\Upsilon$, of a more general quintic NLS equation, which reduces to Eq. (\ref{purequintic}) in an appropriate limit. 
In general, the equivalence group of a differential equation is the group of smooth transformations on the dependent and independent variables that leave the differential structure invariant, whilst simultaneously altering the functional forms of the coefficients. With this in mind, consider the quintic NLS equation
\begin{equation}
\label{quinticlambda}
i\partial_t u + \frac{1}{2m}\partial_{rr} u - \frac{1}{2}\lambda(t)r^2 u - g_0|u|^4 u =0
\end{equation}
where we have introduced the physically 
well-motivated, time-dependent harmonic `trapping' potential, $ \frac{1}{2} \lambda (t) r^2$, 
with $\lambda (t) \in C^{\infty}({\cal{M}})$. The case
$\lambda (t)=0$ corresponds to Eq. (\ref{purequintic}). 
Applying the coordinate transformation 
(\ref{virfinite})-(\ref{virwave}) with $\mu =1/4$ then yields the image of this differential equation \cite{G2001,HU2003,RU2006,UR2008,U2010}:
\begin{equation}
\label{quinticimage}
i\partial_{\tau}\tilde{u} +\frac{1}{2m} \partial_{ss} \tilde{u}- 
\frac{1}{2} \tilde{\lambda} (\tau )s^2 \tilde{u} - g_0 |\tilde{u}|^4 \tilde{u} =0
\end{equation}
where
\begin{equation}
\label{deftildelambda}
\tilde{\lambda} (\tau) := 
(\dot{\beta} (\tau))^{2}  \lambda (\beta (\tau ))  +\frac{m}{2} S[\beta ] (\tau)
\end{equation}
and  
\begin{equation}
\label{schwarzian}
S[\beta ](\tau) := \frac{\dddot{\beta\hspace{0pt}} (\tau)}{\dot{\beta} (\tau)} -\frac{3}{2} \left( \frac{\ddot{\beta} (\tau)}{\dot{\beta} (\tau)} \right)^2
\end{equation}
defines the Schwarzian derivative.  

Hence, Eq. (\ref{quinticlambda}) is form-invariant under 
the group of diffeomorphisms, ${\rm Diff}({\cal{M}})$, represented by Eqs. (\ref{virfinite})-(\ref{virwave}).
Consequently, 
if $u(t,r)$ is a solution to the quintic NLS equation (\ref{quinticlambda}), 
then $\tilde{u} (\tau ,s)$ is a solution to Eqs. (\ref{quinticimage})-(\ref{schwarzian}).
This implies that the equivalence group of Eq. (\ref{quinticlambda}) may be realized as an automorphism on 
the vector space of linear, second-order Schr\"odinger operators on $\mathbb{R}^2$  defined by ${\cal{S}}_{\rm Sch} := \left\{  \Delta_{\rm Sch}\,  \right\}$, where 
\begin{equation}
\label{defSlin}
\Delta_{\rm Sch} := h(t,r) 
\left( i\partial_t +\frac{1}{2m} \partial_{rr} - \frac{1}{2} \lambda (t)r^2 \right) 
\end{equation}
and $h(t,r) \in C^{\infty} (\mathbb{R}^2)$  \cite{U2010}. 
The transformation (\ref{virfinite})-(\ref{virwave}) maps an element in ${\cal{S}}_{\rm Sch}$ onto another element in ${\cal{S}}_{\rm Sch}$. Indeed, we have that 
\begin{equation}
\label{transformD2}
\tilde{\Delta}_{\rm Sch}(\tilde{u}) = (\dot{\beta}(\tau))^{5/4} \exp \left[ -\frac{im  \ddot{\beta}}{4\dot{\beta}} s^2 \right] \Delta_{\rm Sch} (u)
\end{equation}
The representation of the group 
${\rm Diff}({\cal{M}})$ in the space of Schr\"odinger operators ${\cal{S}}_{\rm Sch}$ is then given as 
$\sigma: {\rm Diff}({\cal{M}}) \rightarrow {\rm Hom} ({\cal{S}}_{\rm Sch}, {\cal{S}}_{\rm Sch})$ defined in terms of the left-and-right action \cite{RU2006}
\begin{equation}
\label{leftrightaction}
\sigma (f): \Delta_{\rm Sch} \rightarrow \Pi_{\frac{5}{4}} (f) \circ \Delta_{\rm Sch} \circ \Pi_{\frac{1}{4}} (f^{-1})  
\end{equation}
where $f \in {\rm Diff}({\cal{M}})$ and $\Delta_{\rm Sch} \in {\cal{S}}_{\rm Sch}$. 
The space of Schr\"odinger operators is preserved by the $\sigma$-action of ${\rm Diff}({\cal{M}})$ and may be viewed as a ${\rm Diff}({\cal{M}})$-module. 

The crucial point for the present discussion is that the non-linear quintic term $g_0|u|^4u$ in Eq. (\ref{quinticlambda}) transforms in precisely the same manner as the linear operator, Eq. (\ref{transformD2}), under ${\rm Diff}({\cal{M}})$. Hence, the restriction of $\sigma$ to the affine subspace of operators on the hyperplane $h={\rm constant}$ in ${\cal{S}}_{\rm Sch}$ results in the transformed quintic NLS equation (\ref{quinticimage}). This implies that 
${\rm Diff}({\cal{M}})$ is also the equivalence group for the full class of quintic NLS equations (\ref{quinticlambda}).

\subsection{\label{epsection} On the Space of Sturm-Liouville Operators}

We proceed by considering the linear, second-order Sturm-Liouville (SL) operator:
\begin{equation}
\label{SLoperator}
\Delta_{\rm SL} := m \frac{d^2}{dt^2} +\lambda (t)
\end{equation}
where $\lambda (t) \in C^{\infty}({\cal{M}})$ and $m \in \mathbb{R}$ have the same interpretations discussed above. 
The corresponding Ermakov-Pinney (EP) equation is the non-linear, second-order ODE \cite{E1880,M1930,P1950}
\begin{equation}
\label{NLEP}
\Delta_{\rm SL} (X) = m \frac{d^2X}{dt^2} + \lambda(t) X = \frac{Q}{X^3} , \qquad Q \in \mathbb{R}
\end{equation}
The case $\lambda (t)=0$ is the 
equation of motion for $(0+1)$-dimensional conformal mechanics, $m \ddot{X} =Q/X^3$ \cite{dAFF1976}. 
The Lie point symmetries of this ODE are
\begin{eqnarray}
\label{freeEPisometry}
L_{-1} = -\partial_t , \qquad L_0 & = & -t\partial_t - \frac{1}{2} X\partial_X , 
\nonumber \\
L_1 & = & -t^2\partial_t -tX \partial_X 
\end{eqnarray}
and close under the $sl(2,R)$ algebra. 
The associated infinitesimal transformations may be integrated to the $SL(2,\mathbb{R})$ group action on the plane: 
$t \mapsto (at+b)/(ct+d)$, $X \mapsto X/(ct+d)$, where $ad-bc =1$. 

The generators (\ref{freeEPisometry}) form an $sl(2,R)$ subalgebra of a centreless Virasoro-type algebra ${\rm Vect}({\cal{M}}) := \langle L_n \,  | \, n \in Z \rangle$ spanned by the generators
\begin{equation}
\label{defLn}
L_n = -t^{n+1} \partial_t -\frac{1}{2} (n+1) t^n X \partial_X
\end{equation}
These satisfy the non-vanishing commutation relations $[L_n,L_{n'}] = (n-n')L_{n+n'}$, $\{ n,n' \} \in \mathbb{Z}$.
Eq. (\ref{defLn}) provides a representation of the ${\rm Vect}({\cal{M}})$ algebra as an algebra of differential operators of order one on $\mathbb{R}^2$ spanned by coordinates $(t,X)$. Formal integration of the infinitesimal transformations generated by the operators (\ref{defLn}) yields the finite transformations:
\begin{equation}
\label{EPfinite}
t=\beta (\tau) , \qquad X = (\dot{\beta} (\tau ) )^{1/2} \tilde{X}
\end{equation}

Applying the coordinate transformation (\ref{EPfinite}) maps the EP equation ({\ref{NLEP}) into 
\begin{equation}
\label{transformedEP}
m\frac{d^2\tilde{X}}{d\tau^2} + \tilde{\lambda}(\tau)  =\frac{Q}{\tilde{X}^3}
\end{equation}
where the image of the potential, $\tilde{\lambda}(\tau)$, is given {\em precisely} by Eqs. (\ref{deftildelambda})-(\ref{schwarzian}). The image (\ref{deftildelambda}) was called the `comparison function' by Dingle, who was the first to consider transformations of the form (\ref{EPfinite}) within the context of the EP equation \cite{D1956}.

We conclude, therefore, that the (orientation-preserving) diffeomorphism group ${\rm Diff}({\cal{M}})$ is also the equivalence group of the EP equation (\ref{NLEP}). Indeed, the ${\rm Diff}({\cal{M}})$-actions on the quintic NLS and EP equations are {\em equivalent}, in the sense that the image of the (trapping) potential is the same in both cases \cite{U2010}. 

The form invariance of the EP equation under the ${\rm Diff} ({\cal{M}})$-action (\ref{EPfinite}) admits a geometrical interpretation in terms of the transformation law for tensor-densities. (See, e.g., \cite{O2006,KW2009,O2016} for reviews.) A tensor-density $\alpha$ of degree $h \in \mathbb{R}$ on a one-dimensional manifold ${\cal{M}}$ is defined by
\begin{equation}
\label{deftensor}
\alpha := \alpha (t) (dt)^{h} , \qquad \alpha (t) \in C^{\infty}({\cal{M}})
\end{equation}
Functions on ${\cal{M}}$ have degree $h=0$, vector fields have degree $h=-1$ and quadratic densities have degree $h=2$. The requirement that a tensor-density be coordinate-independent under a diffeomorphism $f: t \mapsto f(t)$ implies that its component should transform as 
\begin{equation}
\label{alphatrans}
f^*_h \cdot \alpha := 
\left( \frac{df^{-1}}{dt} \right)^{h} \cdot \alpha \circ f^{-1}
\end{equation}
 where $f^*$ denotes the pullback to $t$ of $\alpha$ when $\alpha$ is evaluated at $\tau = f(t)$. 
If we define the vector space of tensor-densities of degree $h$ by
${\cal{F}}_{h} ({\cal{M}}) :=\{ \alpha (t)(dt)^h \, | \, \alpha(t) \in C^{\infty}({\cal{M}}) \}$, the transformation 
(\ref{alphatrans}) then represents a $1$-parameter family of ${\rm Diff}({\cal{M}})$-actions on this space. 
In particular, it maps $\Delta_{\rm SL}(X) \mapsto \tilde{\Delta}_{\rm SL} (\tilde{X})$ such that \cite{W1906,C1937}
\begin{eqnarray}
\label{EPimage1}
\tilde{\Delta}_{\rm SL} (\tilde{X}) (df)^{3/2}   & = & \Delta_{\rm SL} (X)  (dt)^{3/2}  \\
\label{EPimage2}
\tilde{X} (df)^{-1/2} & = & X(dt)^{-1/2}
\end{eqnarray}
where, as before, $f := \beta^{-1}$ is the inverse map $\tau = f(t)$.

Hence, it follows from (\ref{EPimage2}) that the argument of the SL operator transforms as a $(-\frac{1}{2})$-density under the ${\rm Diff}({\cal{M}})$-action (\ref{EPfinite}), whereas the image of $X$ under the SL operator, $\Delta_{\rm SL}:X \mapsto \Delta_{\rm SL}(X)$, transforms as a $\frac{3}{2}$-density \cite{DO1994,GO1996}. The SL operator may therefore be interpreted as the map $\Delta_{\rm SL}: {\cal{F}}_{-1/2} \mapsto {\cal{F}}_{3/2}$ from the space of $(-\frac{1}{2})$-densities to the space of $\frac{3}{2}$-densities. Eq. (\ref{alphatrans}) then defines an action of ${\rm Diff}({\cal{M}})$ on the (affine) vector space of linear SL operators ${\cal{S}}_{\rm SL} := \{ \Delta_{\rm SL} \}$. The image of $\Delta_{\rm SL}$ is another SL operator with potential given by (\ref{deftildelambda})-(\ref{schwarzian}). 
We therefore have a representation of ${\rm Diff}({\cal{M}})$ defined by 
$\sigma: {\rm Diff}({\cal{M}}) \rightarrow {\rm Hom} ( {\cal{S}}_{\rm SL},{\cal{S}}_{\rm SL} )$ in terms of the left-and-right action of (\ref{alphatrans}) on the space of SL operators:
\begin{equation}
\label{diffrep}
\sigma (f): \Delta_{\rm LS} \mapsto  f^*_{3/2} \circ \Delta_{\rm SL} \circ (f^*_{-1/2} )^{-1}
\end{equation}
where $f \in {\rm Diff}({\cal{M}})$. 

We now see that the form-invariance of the EP equation under the action (\ref{diffrep}) follows precisely because the non-linear cubic term, $Q/X^3$, transforms under ${\rm Diff}({\cal{M}})$ as a $\frac{3}{2}$-density. Consequently, a solution of the transformed ODE (\ref{transformedEP}) may be generated from a solution to the original equation by specifying $\tilde{X}(\tau) =(\dot{\beta}(\tau))^{-1/2} X(\beta (\tau))$. This is precisely the transformation law for the width of the wavefunction derived from the moment method, as we now discuss in the next Section. 

\renewcommand{\thesection}{3}

\section{\label{momentsection}Moments of The Wavefunction}

Further insight into the dynamics of the non-linear Schr\"odinger equation may be gained by studying the evolution of various integral quantities defined in terms of the wavefunction and its spatial derivatives. These quantities are known as the moments of the wavefunction \cite{T1970,VPT1971,PAB1993,PPV1995,GPT1999,GP1999,G2001}. (See \cite{PTM2005} and references therein.)

Let us consider the $d=1$ NLS equation with an arbitrary non-linear coupling: 
\begin{equation}
\label{genNLS}
i\partial_tu= -\frac{1}{2m} \partial_{rr} u +\frac{1}{2} \lambda (t)r^2 u +g(|u|^2 ,t) u 
\end{equation}
where $g$ is a (possibly explicitly time-dependent) function of the wavefunction modulus. In general, an infinite number of wavefunction moments may be defined \cite{PTM2005}. Those relevant to the present discussion are:
\begin{eqnarray}
\label{mom2}
I_2(t)  & := & m \int dr \, r^2 |u|^2 \\
\label{mom3}
I_3(t) &  := & i\int dr \, r(u\partial_r u^* -u^*\partial_r u) \\
\label{momK}
K(t) & := & \frac{1}{2m} \int dr \, |\partial_r u|^2 \\
\label{momJ}
J(t) & := & \int dr \, G(|u|^2 ,t)
\end{eqnarray}
where in the last expression we have defined $\partial G(|u|^2,t)/\partial |u|^2 := g(|u|^2,t) $. 
Under the assumption that the wavefunction decays at spatial infinity, a 
calculation involving the NLS equation (\ref{genNLS}) and integration by parts reveals that 
\begin{eqnarray}
\label{mom2eom}
\frac{dI_2}{dt} & = & I_3  \\
\label{mom3eom}
\frac{dI_3}{dt} & = & - \frac{2\lambda (t)}{m} I_2 +4K -2 \int dr \, r |u|^2 \frac{\partial g}{\partial r} \\
\label{momKeom}
\frac{dK}{dt} & = & -\frac{\lambda(t)}{2m}  I_3 - 
\frac{i}{2m} \int dr \left( u \frac{\partial u^*}{\partial r} -u^* \frac{\partial u}{\partial r} \right) \frac{\partial g}{\partial r}  \\
\label{momJeom}
\frac{dJ}{dt} & = & \int dr \frac{\partial G}{\partial t} + 
\frac{i}{2m} \int dr \left( u \frac{\partial u^*}{\partial r} -u^* \frac{\partial u}{\partial r} \right) \frac{\partial g}{\partial r}  
\end{eqnarray}

In general, Eqs. (\ref{mom2eom})-(\ref{momJeom}) do not form a closed set of differential equations and moments involving higher-order derivatives of the wavefunction  would be needed to determine the dynamics. This would ultimately necessitate solving an infinite tower of evolution equations. Nonetheless, we may combine Eqs. (\ref{momKeom}) and (\ref{momJeom}) to deduce that 
\begin{equation}
\label{K+Jeom}
\frac{d}{dt} (K+J) =-\frac{\lambda (t)}{2m}  I_3 +\int dr \frac{\partial G}{\partial t}
\end{equation}
Furthermore, we may rewrite Eq. (\ref{mom3eom}) such that 
\begin{equation}
\label{rewritemom3eom}
\frac{dI_3}{dt} =-\frac{2\lambda (t)}{m} I_2 +4 \left[ K +\frac{1}{2} \int dr \left( g|u|^2 -G \right) \right]
\end{equation}
Comparison between Eqs. (\ref{K+Jeom}) and (\ref{rewritemom3eom}) then implies that necessary and sufficient conditions for closure of the system (\ref{mom2eom})-(\ref{momJeom}) are that
\begin{equation}
\label{neccond}
J=\frac{1}{2} \int dr \left( g|u|^2 -G \right)
\end{equation}
and that the integral $\int dr (\partial G/\partial t)$ be directly expressible only in terms of the dependent functions \cite{PTM2005}. Given the definition of the moment $J$ in Eq. (\ref{momJ}), condition (\ref{neccond}) reduces to the requirement that 
\begin{equation}
\label{requirement}
\int dr (3G -g|u|^2) =0
\end{equation}
On the other hand, since (by assumption)  $G$ does not depend explicitly on the spatial coordinate, this implies that the integrand must vanish identically. This in turn implies after integration that the coupling must take the form 
\begin{equation}
\label{neccond1}
g(|u|^2,t)=g_0(t)|u|^4 , \qquad g_0(t) \in C^{\infty}(\mathbb{R})
\end{equation}
However, in this case, the second necessary condition for closure reduces to the requirement that  
\begin{equation}
\label{neccond2}
\int dr \frac{\partial G}{\partial t} = \frac{1}{g_0} \frac{dg_0}{dt} J
\end{equation}
and closure therefore requires that $g_0$ be constant in time for this term to cancel when $J \ne 0$ \cite{PTM2005}.  

In summary, therefore, the non-linear coupling in the Schr\"odinger equation (\ref{genNLS})
must satisfy the condition $g(|u|^2 ,t) =g_0 |u|^4$, $g_0 \in \mathbb{C}$ if the (lowest-order) wavefunction moments are to form a closed system of ODEs.  
This is {\em precisely} the condition for a generic quintic NLS equation to admit a six-dimensional Lie point symmetry which is isomorphic to the Lie algebra $sch_1$ \cite{OG2012}. 

The corresponding non-autonomous, linear system of wavefunction moment equations is therefore 
\begin{eqnarray}
\label{closed2}
\frac{dI_2}{dt} & = & I_3 \\
\label{closed3}
\frac{dI_3}{dt} & = & -\frac{2 \lambda(t)}{m}  I_2 +4I_4 \\
\label{closedham}
\frac{dI_4}{dt} & = &
-\frac{\lambda (t)}{2m} I_3
\end{eqnarray}
where we have defined $I_4 := K+J$. A physically intuitive interpretation of the wavefunction moments is given in terms of the square of the width $(I_2)$, the momentum $(I_3)$ and the energy $(I_4)$ of the wavefunction, respectively. Note that the effects of the non-linear term in 
the Schr\"odinger equation are entirely encoded within the energy functional $I_4$. The form of the system (\ref{closed2})-(\ref{closedham}) remains invariant in the limit $g_0 \rightarrow 0$ $(J\rightarrow 0)$, corresponding to the free Schr\"odinger equation. 

The evolution equations (\ref{closed2})-(\ref{closedham}) admit a constant of motion: 
\begin{equation}
\label{defQ}
\frac{Q}{m}  := 2I_2 I_4 - \frac{1}{4} I^2_3
\end{equation}
which is independent of the trapping potential. (We identify this parameter as a separation constant in Section \ref{similarsection}.) A short calculation then reveals that the width of the wavefunction, $X := I_2^{1/2}$, satisfies the EP equation (\ref{NLEP}) \cite{VPT1971}:
\begin{equation}
\label{EP}
\Delta_{\rm SL} (X) = m\frac{d^2X}{dt^2} +\lambda (t)  X =\frac{Q}{X^3}
\end{equation}
System (\ref{closed2})-(\ref{closedham}) can therefore be solved in its entirety given a solution to (\ref{EP}). But the general solution to (\ref{EP}) is given by a non-linear superposition \cite{P1950}
\begin{equation}
\label{EPgensol}
X_{\rm EP} =\left[ AX_1^2 +BX_2^2 +2C X_1X_2 \right]^{1/2}
\end{equation}
where $(X_1(t),X_2(t))$ are two linearly independent solutions to the corresponding Sturm-Liouville  ODE, $\Delta_{\rm SL} (X)=0$. The constants $A,B,C \in \mathbb{R}$ satisfy the constraint $AB-C^2 =Q/(mW^2)$, where the constant $W:= X_1\dot{X}_2 - X_2\dot{X}_1$ is the Wronksian. Hence, the problem of solving the moment equations (\ref{closed2})-(\ref{closedham}) is reduced to finding two linearly-independent solutions to the SL equation. 

An important observation to make is that closure of (\ref{mom2eom})-(\ref{momJeom}) is possible for an {\em arbitrary} time-dependence in the quadratic trapping potential, $\frac{1}{2}\lambda (t)r^2$. On the other hand, the discussion of Section \ref{schsection} has shown that the (affine) vector space of Schr\"odinger operators (\ref{defSlin}) is a ${\rm Diff}({\cal{M}})$-module under the ${\rm Diff}({\cal{M}})$-action (\ref{virfinite})-(\ref{virwave}), where the trapping potential is mapped to its image (\ref{deftildelambda})-(\ref{schwarzian}). We may immediately deduce, therefore, that the system of moment equations (\ref{closed2})-(\ref{closedham}) is also form-invariant under the ${\rm Diff}({\cal{M}})$-action (\ref{virfinite})-(\ref{virwave}), with the image of the trapping potential given by Eqs. (\ref{deftildelambda})-(\ref{schwarzian}). The moments transform such that $I_j \mapsto \tilde{I}_j$, where tilde quantities are defined as before, but now in terms of the wavefunction $\tilde{u}$ and coordinates $(\tau ,s)$. In particular, the wavefunction width transforms as
$\tilde{X}(\tau ) = (\dot{\beta}(\tau ))^{-1/2} X(\beta( \tau) )$, which is precisely 
the  $(-\frac{1}{2})$-density transformation given in Eq. (\ref{EPimage2}) for the argument of the SL operator in the EP equation. This equivalence between the ${\rm Diff}({\cal{M}})$ actions on the Schr\"odinger and Sturm-Liouville operator spaces is the reason why the moment method is able to (partially) preserve the dynamics of the quintic NLS equation in terms of the EP equation for the wavefunction width. The connection between the two can be traced back to the underlying $SL(2,\mathbb{R})$ kinematical symmetry in the absence of a trapping potential. 

In the following Section, we exploit this equivalence within a cosmological context. 

\renewcommand{\thesection}{4}

\section{\label{cosmosection} Cosmology of the Moment}

\subsection{\label{frwsection} Cosmological Friedmann Equations}

For the quintic Schr\"odinger equation (\ref{purequintic}) in the absence of a trapping potential, the three conserved quantities associated with the generators of the $SL(2,\mathbb{R}) \cong SO(2,1)$ kinematical symmetry (\ref{sch1rep})
are the Hamiltonian, ${\cal{I}}_4$, the dilatation, ${\cal{D}}$, and the conformal generator, ${\cal{K}}$, respectively. These are given by 
\begin{eqnarray}
\label{conservedham}
{\cal{I}}_4 = \int dr \left[ \frac{1}{2m} |\partial_r u|^2 +g_0 |u|^6 \right]
\\
\label{conserveddil}
{\cal{D}} =t{\cal{I}}_4 -\frac{1}{4}{\cal{I}}_3
\\
\label{conservedconf}
{\cal{K}} = -t^2{\cal{I}}_4 +2t{\cal{D}} +\frac{1}{2}{\cal{I}}_2
\end{eqnarray}
where the moments ${\cal{I}}_2$, ${\cal{I}}_3$ are defined in (\ref{mom2})-(\ref{mom3}). By Noether's theorem, these generators are constants in time, so 
\begin{equation}
\label{emptywave}
\frac{d{\cal{I}}_2}{dt}={\cal{I}}_3, \qquad \frac{d{\cal{I}}_3}{dt} =4 {\cal{I}}_4 , 
\qquad \frac{d{\cal{I}}_4}{dt}=0
\end{equation}
which implies that 
\begin{equation}
\label{emptyEP}
m\frac{d^2{\cal{X}}}{dt^2} = \frac{Q}{{\cal{X}}^3}, \qquad \frac{Q}{m}:= 2{\cal{I}}_2{\cal{I}}_4 -\frac{1}{4}{\cal{I}}_3^2 , \qquad \frac{dQ}{dt} =0
\end{equation}
where ${\cal{X}} := {\cal{I}}_2^{1/2}$. The constant of motion, $Q$, is the quadratic Casimir operator of the $SL(2,\mathbb{R})$ symmetry and commutes with all three generators 
$({\cal{I}}_4, {\cal{D}}, {\cal{K}} )$ with respect to Poisson brackets that are closed under the $sl(2,\mathbb{R})$ algebra. 

In general, the $SL(2,\mathbb{R})$ symmetry of (\ref{emptywave})-(\ref{emptyEP}) is explicitly broken by an arbitrary time-reparametrization under the  ${\rm Diff}({\cal{M}})$-action (\ref{virfinite}). This action generates a trapping potential via Eq. (\ref{deftildelambda}) and results in the set of moment equations (\ref{closed2})-(\ref{EP}), where the potential is given in terms of the Schwarzian derivative  (we drop tildes in what follows to ease notation):
\begin{equation}
\label{lambdaschwarzian}
\lambda (\tau) =\frac{m}{2} S[\beta ](\tau) = \frac{m}{2} \left(
\frac{\dddot{\beta\hspace{0pt}} (\tau)}{\dot{\beta} (\tau)} -\frac{3}{2} \left( \frac{\ddot{\beta} (\tau)}{\dot{\beta}(\tau)} \right)^2 \right)
\end{equation}
The kernel of the Schwarzian map is the group of fractional linear transformations $t \mapsto (at+b)/(ct+d)$, $ad-bc=1$, so it is the elements $\beta^{-1} (t) \in {\rm Diff}({\cal{M}})$ which are not in this kernel that explicitly break the $SL(2,\mathbb{R})$ symmetry and generate a non-trivial trapping potential from the vanishing case, $\lambda (t) =0$. 

We are now ready to uncover the hidden cosmological dynamics. We define functions $a(\tau), \phi (\tau) ,\rho (\tau)$, a new time parameter $T$ and a constant $G \in \mathbb{R}$ such that 
\begin{eqnarray}
\label{defphi}
8\pi G \left( \frac{d\phi}{d\tau} \right)^2 := S[\beta ](\tau) \\
\label{defrho}
\rho (\tau) := \frac{3}{4\pi G} I_4 \\
\label{defa}
a(\tau ) := X(\tau) = I_2^{1/2}\\
\label{defT}
\frac{d}{dT} := a \frac{d}{d\tau}
\end{eqnarray}
Eqs. (\ref{closed2})-(\ref{EP}) then transform to 
\begin{eqnarray}
\label{frw1}
H^2 = \frac{\dot{a}^2}{a^2} =\frac{8\pi G}{3} \rho -\frac{k}{a^2}
\\
\label{frw2}
\dot{\rho} =-3H\dot{\phi}^2 
\\
\label{frw3}
\frac{\ddot{a}}{a}-\frac{\dot{a}^2}{a^2} -\frac{k}{a^2} =-4\pi G \dot{\phi}^2 
\end{eqnarray}
where we have relabeled $Q/m :=k$ and a dot denotes differentiation with respect to $T$ \cite{L2003,L2013}. We recognize Eqs. (\ref{frw1})-(\ref{frw3}) as the field equations of Einstein gravity for a spatially isotropic and homogeneous Friedmann-Robertson-Walker (FRW) cosmology sourced by a minimally coupled, self-interacting scalar field $\phi$ with energy density $\rho$. We identify $G$ as Newton's constant and $H(T) := \dot{a}/a$ as the Hubble expansion parameter. The line-element of the corresponding FRW spacetime is 
\begin{eqnarray}
\label{lineelement}
ds^2 & = &  -N^2(\tau)d\tau^2 +a^2(\tau) d\Omega^2_{3,k} 
\nonumber \\
& = & -dT^2 +a^2(T) \left( \frac{dx^2}{1-kx^2} +x^2d\Omega_2^2 \right)
\end{eqnarray}
where $T$ denotes cosmic time, the width of the wavefunction is interpreted directly as the scale factor of the universe $X=a$ and the spatial three-sections $d\Omega^2_{3,k}$ have constant curvature $k \in \{ 0,\pm 1 \}$. The metric may also be expressed in terms of the lapse function, $N(\tau) =1/a(\tau)$, and the time coordinate $\tau$.

Under the correspondence (\ref{defphi})-(\ref{defT}), the inverse of the wavefunction momentum transforms via Eq. (\ref{closed2}) into the comoving Hubble radius, $2I_3^{-1} = 1/(aH)$. This determines an effective scale for the cosmic horizon during an epoch of inflationary expansion. Perturbations that are generated quantum-mechanically during inflation are effectively frozen when their wavelength exceeds this length-scale. 

Eq. (\ref{frw1}) is the Friedmann equation. This equation follows as a direct consequence of the definition of the constant of motion $Q$ in Eq. (\ref{defQ}). 

Eq. (\ref{frw2}) is the equation of energy-momentum conservation for the scalar field. This equation follows most directly from the evolution equation for the Hamiltonian of the wavefunction, Eq. (\ref{closedham}). The self-interaction potential of the scalar field, $V(\phi)$, is defined through 
$\rho := \frac{1}{2}\dot{\phi}^2 +V(\phi)$. 

Eq. (\ref{frw3}) follows from transforming the EP equation (\ref{EP}). The interpretation of the transformed EP equation from a gravitational perspective is somewhat ambiguous. Within the context of standard General Relativity, the left-hand side of Eq. (\ref{frw3}) is (one half) the sum of the time-time and space-space components of the Einstein field equations (EFEs). The right-hand side can be interpreted as the inertial mass of the matter source. The Friedmann equation is the time-time component of the EFEs and Eqs. (\ref{frw1}) and (\ref{frw3}) together constitute the full set of non-trivial components of the EFEs when restricted to the FRW metric. Given these two equations, conservation of energy-momentum follows automatically from the Bianchi identity. Thus, only two of the three equations (\ref{frw1})-(\ref{frw3}) are independent. (This also implies that  Eq. (\ref{closed3}) does not introduce a further dynamical constraint into the system.)

On the other hand, in unimodular versions of gravity, where a constraint is imposed on the determinant of the metric, the cosmological equations are given by the traceless Einstein field equations \cite{E1919,U1989,W1989,FGB2000}: 
\begin{equation}
\label{traceless}
R_{\mu\nu} -\frac{1}{4}Rg_{\mu\nu} =8\pi G \left( T_{\mu\nu} - \frac{1}{4} Tg_{\mu\nu} \right)
\end{equation}
For FRW metrics, all non-trivial components of the traceless field equations reduce to the single equation (\ref{frw3}). However, conservation of energy-momentum no longer follows from the Bianchi identity and must be imposed by hand. If energy-momentum conservation is imposed, the Friedmann equation (\ref{frw1}) then represents a first integral of the system \cite{W1989,EEMU2011,E2013}.

Within the context of the present discussion, the Casimir constraint equation (\ref{defQ}) is a first integral of the system of moment equations. The origin of the constant, $Q$, lies in the underlying kinematical $SL(2,\mathbb{R})$ symmetry exhibited by the constant-Hamiltonian system of equations (\ref{emptywave})-(\ref{emptyEP}). Indeed, Q represents the quadratic Casimir operator of the $SL(2,\mathbb{R})$ symmetry in this case. Since the space of Schr\"odinger operators (\ref{defSlin}) is a ${\rm Diff}({\cal{M}})$-module under the action of Eqs. (\ref{virfinite})-(\ref{virwave}), an integral constraint also arises in the transformed system. Remarkably, this constraint corresponds precisely to that of a first integral for gravitational field equations that respect conservation of energy-momentum.  In this sense, energy-momentum conservation has followed naturally from the dynamics of the wavefunction moments and the underlying $SL(2, \mathbb{R})$ symmetry. It seems more natural therefore to interpret the constraint equation (\ref{frw1}) as a genuine  first integral of the field equations rather than as a separate component, and to then regard Eqs. (\ref{frw1})-(\ref{frw3}) in terms of the traceless Einstein equations for the FRW metric. In this case, the integration constant $\Lambda \in \mathbb{R}$ plays the role of a cosmological constant and can be absorbed into the definition of the total energy density $\rho \mapsto \rho_{\phi} + \Lambda$, where $\rho_{\phi}$ now denotes the scalar field energy density.

\subsection{\label{orbitsection} Cosmic Dynamics and the Orbit of de Sitter Cosmology}

From the wavefunction perspective, the $SL(2,\mathbb{R})$ symmetric, zero energy configuration is ${\cal{I}}_4 =0$. This has general solution satisfying ${\cal{I}}_3^2=-4Q$. The configuration is static (constant width) for $Q=0$, which just corresponds to empty Minkowski spacetime. If $Q<0$, we have that $\dot{a}^2(T) =1$ in a negatively-curved universe. This is the Milne universe describing a quarter wedge of Minkowski spacetime expressed in expanding coordinates \cite{M1935}. 

The general $SL(2,\mathbb{R})$ symmetric wavefunction with a non-zero, constant Hamiltonian has width (squared) determined by the EP equation (\ref{emptyEP}). The general solution is given by Eq. (\ref{EPgensol}):
\begin{equation}
\label{genempty}
{\cal{X}}^2 (t) = {\cal{I}}_2(t) =(A+Bt)^2 +\frac{Q}{mA^2} t^2
\end{equation}
where $A,B \in \mathbb{R}$ are arbitrary integration constants. This solution corresponds cosmologically to the de Sitter universe sourced by a spacetime-independent cosmological constant. For example, consider the spatially flat cosmology, $Q=k=0$. From (\ref{frw1}) we deduce the exponential expansion $a=a_i \exp [ B(T-T_i)]$, $a_i,T_i \in \mathbb{R}$. Note that the cosmological constant arises here as an integration constant, as would be expected for the traceless EFEs. It is important to note that given a particular solution to the EP equation (\ref{emptyEP}), the general solution (\ref{genempty}) is generated by the $SL(2,\mathbb{R})$ transformation: $t\mapsto (\alpha t+\beta)/(\gamma t+\delta)$, ${\cal{X}} \mapsto (\gamma t+\delta ){\cal{X}} $, $\alpha \delta - \beta \gamma =1$. In this sense, the Minkowski and de Sitter spacetimes should be regarded as elements of the same $SL(2,\mathbb{R})$ equivalence class of solutions characterized by different choices of initial conditions (integration constants). This is important in what follows. 

The vacuum solution (\ref{genempty}) may be employed as the seed for generating dynamical scalar field cosmologies via the ${\rm Diff}({\cal{M}})$-action (\ref{virfinite})-(\ref{virwave}) (or equivalently Eq. (\ref{EPfinite})). Here, we focus on the spatially flat $(k=0)$ models. The non-linear superposition property (\ref{EPgensol}) of the EP equation implies that the dynamics of spatially curved models may be written down directly from the linearly independent solutions to the flat model \cite{HL2002}. We normalize units such that $8 \pi G=1$.

The ${\rm Diff}({\cal{M}})$-action (\ref{EPfinite}), $t=\beta (\tau)$, together with the $(-\frac{1}{2})$-density 
transformation of the wavefunction width, $X = (\dot{\beta} (\tau))^{-1/2}{\cal{X}}$, generates the cosmic scale factor
\begin{equation}
\label{diffa}
a(\tau ) =\frac{A +B\beta(\tau)}{\sqrt{\dot{\beta} (\tau )}}
\end{equation}
The associated scalar field evolution is determined by formally integrating the definition (\ref{defphi}):
\begin{equation}
\label{diffphi}
\phi (\tau) =\int d\tau \, \sqrt{S[ \beta ](\tau)}
\end{equation}
The self-interaction potential of the scalar field is then determined directly from the first-integral constraint (\ref{frw1}):
\begin{equation}
\label{diffpot}
V(\tau) =3 (\dot{a}(\tau))^2 -\frac{1}{2} a^2(\tau) S[\beta ](\tau)
\end{equation}
Cosmic time is determined parametrically in terms of the quadrature:
\begin{equation}
\label{diffT}
T(\tau ) =\int \frac{d\tau}{a(\tau)} = \int d\tau \frac{ (\dot{\beta}(\tau))^{1/2}}{A+B\beta (\tau)}
\end{equation}

Thus, we arrive at a parametric scalar field cosmology in terms of the dynamical variable $\tau$. Depending on the particular element of ${\rm Diff}({\cal{M}})$, i.e., the functional form of $\beta (\tau)$, we may express the solution in terms of cosmic time by integrating (\ref{diffT}) and inverting the result. 

Each map $\beta (\tau) \in {\rm Diff}({\cal{M}})$ generates a rolling scalar field and this suggests that a given cosmology may be identified with an element of ${\rm Diff}({\cal{M}})$. Alternatively, 
we are effectively associating each dynamical scalar field cosmology with a non-trivial trapping potential, $\lambda (t) \ne 0$, corresponding to a `point' in the (affine) vector space of Schr\"odinger operators defined by Eq. (\ref{defSlin}). Different points are reached from $\lambda (t) = 0$ by the representation of ${\rm Diff}({\cal{M}})$ given by Eq. (\ref{leftrightaction}). However, it is important to emphasize that the non-trivial Schwarzian derivative term in the image of the potential, Eq. (\ref{deftildelambda}), is proportional to the mass parameter, $m$. It is therefore more accurate to parametrize each element in the space of operators ${\cal{S}}_{\rm Sch}$ in terms of the components $(\lambda, m)$ of a vector. (We shall see in the next Section that under appropriate restrictions, $m$ is proportional to the central charge of the Virasoro group.) Crucially, the transformation from 
$(0,m) \mapsto (\frac{m}{2} S[\beta](\tau) , m )$ is non-trivial precisely because $m \ne 0$. 

Thus, each scalar field cosmology $(\frac{m}{2} S[\beta](\tau) , m  )$
can be reached from the `de Sitter' point $(0,m)$  by the ${\rm Diff}({\cal{M}})$-action (\ref{leftrightaction}). We associate this de Sitter point with the identity element $t=f^{-1}(\tau) = \beta(\tau) =\tau$ in ${\rm Diff}({\cal{M}})$. 
However, one must also take into consideration the underlying $SL(2,\mathbb{R})$ symmetry by determining the stability subgroup of this element. In general, the action of a Lie group $G$ (with Lie algebra $\mathbf{g}$) on a manifold ${\cal{M}}$ is the smooth map $G\times {\cal{M}} \rightarrow {\cal{M}}: (g,x) \mapsto g(x)$, where $g \in G$, $x \in {\cal{M}}$. The orbit of an element $x \in {\cal{M}}$ is the set of elements in ${\cal{M}}$ that can be reached from $x$ by the group action, ${\cal{O}}_x := \{ g(x) \, | \, g\in G \}$. There is a bijection between the orbit and the quotient manifold ${\cal{O}}_x \cong G/{\rm Stab}_x$, where ${\rm Stab}_x := \{ g \in G \, | \, g(x)=x\}$ is the stabilizer of $x$ \cite{K2004}. Each smooth action of $G$ on ${\cal{M}}$ induces a corresponding Lie algebra action  $Z(x)$, $Z \in \mathbf{g}$. The subalgebra ${\rm stab}_x := \{ Z \in \mathbf{g} | Z(x) =0 \} \subset \mathbf{g}$ is the stabilizer algebra of $x$. It is straightforward to prove that the stabilizer algebra ${\rm stab}_x$ is the Lie algebra of ${\rm Stab}_x$ \cite{K2004}. 

We must therefore determine the stabilizer of the ${\rm Diff}({\cal{M}})$-action (\ref{leftrightaction}) on the space of Schr\"odinger operators (\ref{defSlin}). Since this representation is equivalent to the action 
(\ref{diffrep}) on the space of SL operators (\ref{SLoperator}), let us define the stability subgroup of 
${\cal{S}}_{\rm SL} = \{ \Delta_{\rm SL} \}$ in ${\rm Diff}({\cal{M}})$ to be  
${\rm Stab}_{\lambda} := \{ f \in {\rm Diff}({\cal{M}})| f_*(\Delta_{\rm SL} ) = \Delta_{\rm SL} \}$, 
where $f_*$ is the pushforward of the ${\rm Diff}({\cal{M}})$ action on $\Delta_{\rm SL}$. 
Now consider the infinitesimal diffeomorphism $\beta(\tau):= \tau + \epsilon v(\tau)$ for a `small' parameter $\epsilon$ and $v (\tau) \in C^{\infty}({\cal{M}})$. Then the infinitesimal action of ${\rm Diff}({\cal{M}})$ on the solutions to the ODE, $\Delta_{\rm SL}(X)=0$, is given by 
\begin{eqnarray}
\label{infinX}
\tilde{X} & = & X(\tau) +\epsilon \left( v \partial_{\tau} X-\frac{1}{2} X \partial_{\tau}v \right) \nonumber \\\tilde{\lambda} & = & \lambda (\tau) +\epsilon \left( v \partial_{\tau} \lambda +2\lambda \partial_{\tau} v +\frac{m}{2}\partial^3_{\tau} v \right)
\end{eqnarray}
where the second equation is the infinitesimal version of (\ref{deftildelambda}). (Note that the third derivative term arises from the infinitesimal limit of the Schwarzian derivative.) It then follows from Eq. (\ref{infinX}) that $v \in {\rm Lie}({\rm Stab}_{\lambda})$ if and only if
\begin{equation}
\label{stab}
v \partial_{\tau} \lambda +2\lambda \partial_{\tau} v +\frac{m}{2} \partial^3_{\tau} v =0
\end{equation}

For the de Sitter cosmology $(\lambda,m)=  (0,m)$, the stabilizer is $\partial^3_{\tau} v =0$. The general form
of the $v(\tau)$-expansion is therefore $v(\tau)= a+b\tau+c\tau^2$, where the arbitrary constants $a,b,c \in \mathbb{R}$ can be associated with the time translation $(\delta \tau =\mu)$, dilatation $(\delta \tau = \mu \tau)$ and conformal boost $(\delta \tau= \mu \tau^2)$
generators of the $sl(2,\mathbb{R})$ algebra, respectively. Hence, the vector field $V=v\partial_{\tau}$ generates the $sl(2,\mathbb{R})$ algebra, so the stabilizer of $(0,m)$ is the group $SL(2,\mathbb{R})$. Indeed, as we have discussed above, this also follows directly from Eq. (\ref{deftildelambda}), since the kernel of the Schwarzian operator, $S: t\mapsto S[f](t)$, is the group of $SL(2,\mathbb{R})$ transformations $f: t \mapsto  (at+b)/(ct+d)$, $ad-bc=1$. 

We conclude, therefore, that each scalar field cosmology may be identified 
with an element on the orbit of the de Sitter point $(0,m)$ generated under the orientation-preserving ${\rm Diff}({\cal{M}})$-action. This orbit is the quotient manifold 
\begin{equation}
\label{orbitcosmo}
{\cal{O}}_{(0,m)} = \frac{{\rm Diff^+} (\mathbb{R})}{SL(2,\mathbb{R})}
\end{equation}
Consequently, we may label each scalar field cosmology with a $\beta (\tau)$, where it is understood that we mod out by an $SL(2,\mathbb{R})$ transformation. In other words, we have the equivalence relation $\beta (\tau) \sim h(\beta (\tau ))$, where $h(\tau) = (l\tau +m)/(n \tau + p)$, $lp-mn =1$. The identity element on the orbit of $(0,m)$ is $\beta =\tau \, {\rm mod} \, SL(2,\mathbb{R})$. Maps of the form $\beta (\tau) =\tau +\epsilon v(\tau )$ that perturb away from the identity may be viewed as points along the orbit in the neighbourhood of the exponentially expanding de Sitter spacetime. Since such maps generate a dynamical scalar field, these will correspond to models of slow-roll inflation. 

\subsection{\label{similarsection} Cosmology and Self-Similar Wavefunctions}

An important question to address at this stage is the nature of the wavefunctions 
that correspond to the FRW scalar field cosmologies under consideration. 
Here, we show that the appropriate wavefunctions are self-similar. In general, 
self-similar wavefunctions have the form 
$u = A(t)\chi [r/b(t)] e^{i\alpha(t,r)}$, where $b(t)$ is interpreted physically as the width of the wavefunction
\cite{RVLTB2000,P2003}. It has been shown that all finite energy, self-similar solutions to Eq. (\ref{genNLS}) have a phase that is necessarily quadratic in the spatial coordinate with a time-dependent coefficient (referred to as the chirp) proportional to the logarithmic derivative of the width, $\alpha (t,r)= (m\dot{b}/2b)r^2$ \cite{P2003}. Indeed, the self-similar solution derived from the quintic NLS equation (\ref{quinticlambda}) is given by \cite{KNSX2000}
\begin{equation}
\label{similarquintic}
u =\frac{1}{b^{1/2} (t)} \chi (\sigma) \exp \left[ \frac{im}{2}\frac{\dot{b}}{b} r^2 \right]
\end{equation}
where $\sigma := r/b(t)$ and the functions $b(t)$ and $\chi (\sigma )$ satisfy the equations
\begin{eqnarray}
\label{similar1}
m \ddot{b} +\lambda (t)b =\frac{Q}{b^3} \\
\label{similar2}
\chi_{\sigma \sigma} =mQ\chi \sigma^2 +2mg_0\chi^5
\end{eqnarray}
respectively, where $Q \in \mathbb{R}$ arises as an arbitrary separation constant. It has been further shown that a quadratic phase in the wavefunction is precisely the condition required to lead to a closed system of moment equations (\ref{closed2})-(\ref{closedham}) \cite{PTM2005}. In this case, consistency with the self-similar solution (\ref{similarquintic}) and the moment equation (\ref{closed2}) implies that $I_2=b^2=X^2$, in agreement with our interpretation of $X$ as the wavefunction width.  

The scalar field cosmologies we have considered above may therefore be associated with exact self-similar solutions of the quintic NLS equation (\ref{quinticlambda}). This is a natural association given that a characteristic feature of self-similar solutions is that they should exhibit no changes in shape when evolving, in line with the notion of an isotropically expanding (and/or contracting) cosmology that remains homogeneous. 

This association may be made more concrete by re-examining the transformed wavefunction (\ref{virwave}) under the action of ${\rm Diff}({\cal{M}})$:
\begin{equation}
\label{similarimage}
\tilde{u}(\tau ,s) = ( \dot{\beta}(\tau ) )^{1/4} \exp \left[ -\frac{im}{4} 
\frac{\ddot{\beta}}{\dot{\beta}} s^2 \right] \hat{u} (t,r)
\end{equation}
Comparison  between Eqs. (\ref{similarquintic}) and (\ref{similarimage})  then implies that we may interpret solution (\ref{similarquintic}) as the  self-similar wavefunction generated from the constant Hamiltonian solution $\hat{u}$. This latter wavefunction is determined by solving the corresponding equations (\ref{similar1})-(\ref{similar2}) for a zero trapping potential $\lambda (t)=0$. It has a width $\hat{b} =A+Bt$ (when $Q=0$) and time-independent quadratic phase. Hence, the width of the transformed wavefunction (\ref{similarimage}) is given by $\tilde{b}(\tau ) = (A+B\beta (\tau))/\dot{\beta}^{1/2}$, in agreement with the expression for the cosmic scale factor, Eq. (\ref{diffa}). (Note that the independent variable $\sigma$ transforms as a singlet under the ${\rm Diff}({\cal{M}})$-action (\ref{virfinite})-(\ref{virwave}), so the functional form of $\chi (\sigma)$ remains invariant. Likewise, the relationship between the amplitude and chirp of the wavefunction, as defined above in terms of the wavefunction width, is also respected.  
The self-similar character of the wavefunction is therefore preserved under the ${\rm Diff}({\cal{M}})$-action.)   

\subsection{\label{examplesection} Power Law Diffeomorphisms and Exponential Potential Models}

Now that we have established the role played by the $SL(2,\mathbb{R})$ symmetry in governing the dynamics of FRW scalar field cosmologies, we illustrate this with a worked example. Given the form of the non-trivial (traceless) EFE in terms of the EP equation, together with the invariance of the Schwarzian derivative under the $SL(2,\mathbb{R})$ diffeomorphism,
it is natural to identify scalar field cosmologies through the functional form of the 
field's kinetic energy, $K[\phi(\tau)] := (d\phi/d\tau)^2$, parametrized with respect to the time coordinate, $\tau$. This is manifestly $SL(2,\mathbb{R})$ invariant. Consequently, for a given $K[\phi (\tau)]$, the general dependence of the scale factor on $\tau$ remains unaltered by an $SL(2,\mathbb{R})$ time reparametrization (up to linear translations on the numerical values of the (arbitrary) integration constants $A$ and $B$ in Eq. (\ref{diffa})). Different particular solutions within an $SL(2,\mathbb{R})$-equivalence class then correspond to different choices of these integration constants, i.e., to different choices of initial conditions. 
For rolling scalar fields, two particular solutions within an $SL(2,\mathbb{R})$ equivalence can be mapped onto one another by means of a particular $SL(2,\mathbb{R})$ transformation. The scalar field will exhibit the same $\tau$-dependence in both cases. Each particular solution would then correspond to a different self-interaction potential for the scalar field, and consequently, would originate from different underlying particle physics theories. In this sense, different particle physics models can be viewed as being $SL(2,\mathbb{R})$-equivalent. 

By way of illustration, consider the power-law diffeomorphism:
\begin{equation}
\label{power}
t= \beta (\tau ) = \tau^n , \qquad n \in \mathbb{R} \backslash \{\pm 1\}
\end{equation}
This yields a Schwarzian 
\begin{equation}
\label{powersch}
S[\beta](\tau) = \left( \frac{1-n^2}{2} \right) \frac{1}{\tau^2}
\end{equation}
and, from Eq. (\ref{diffphi}), a corresponding scalar field evolution 
\begin{equation}
\label{powerphi}
\phi(\tau) -\phi_0  =\sqrt{\frac{1-n^2}{2}} \ln \tau
\end{equation}
where $\phi_0 \in \mathbb{R}$ is an arbitrary integration constant. Note that $n^2 <1$ is necessary for the scalar field to have positive-definite kinetic energy. 

The two fundamental solutions to Eq. (\ref{emptyEP}) when $Q=0$ are ${\cal{X}}_1 =Bt$ and ${\cal{X}}_2=A$, where $A,B\in \mathbb{R}$. From these we generate the scale factors and scalar field self-interaction potentials from Eqs. (\ref{diffa})-(\ref{diffT}): 
\begin{eqnarray}
\label{power2}
a_1 & \sim & \tau^{(n+1)/2} \sim T^{(1+n)/(1-n)} , 
\nonumber \\
V_1 (\phi) & = & V_{1,0} (1+2n) \exp \left[ - \sqrt{\frac{2(1-n)}{1+n}} \phi \right]
\end{eqnarray}
and 
\begin{eqnarray}
\label{power1}
a_2 & \sim & \tau^{(1-n)/2} \sim T^{(1-n)/(1+n)} ,
\nonumber \\
V_2(\phi) & = & V_{2,0} (1-2n) \exp \left[ -\sqrt{\frac{2(1+n)}{1-n}} \phi \right]
\end{eqnarray}
respectively, where $V_{i,0}$ are arbitrary, positive-definite constants. These two solutions are related via the map $n \mapsto -n$. In terms of cosmic time, the power of the expansion $a \sim T^p$ is inverted such that $p \mapsto 1/p$. Hence, the map transforms an accelerating, inflationary cosmology into a decelerating, deflationary solution, and vice-versa. The invariant background is the coasting cosmology $a \sim T$. Note that for $|n|>1/2$, the potentials have opposite sign, so will in general originate from differing underlying particle physics theories. When $n>1/2$, for example, the time-reversal of Eq. (\ref{power1}) corresponds to the decelerating, contracting solution which forms the basis behind the cyclic scenario for the early universe \cite{BST2004}. This is `dual' to the power-law inflationary cosmology (\ref{power2}) \cite{L2004}. Indeed, the two scenarios are related by the $SL(2,\mathbb{R})$ reparametrization $\tau \mapsto -1/\tau$, which is one of the generators of the $SL(2,\mathbb{R})$ group. 

The general solution to the (traceless) EFE (\ref{frw3}) in parametric form is deduced directly from the linear combination of particular solutions (\ref{power2})-(\ref{power1}), or equivalently, from the full expression (\ref{diffa}):
\begin{equation}
\label{powergen}
a(\tau) = a_{1,0} \tau^{(1+n)/2} +a_{2,0} \tau^{(1-n)/2}
\end{equation}
where $a_{1,0}$ and $a_{2,0}$ are arbitrary, positive constants. The integral in Eq. (\ref{diffT}) can be evaluated analytically in terms of a Gauss hypergeometric function, but the result is not in general invertible:
\begin{eqnarray}
\label{powerhyper}
T & = &  \frac{2}{a_{2,0}(1+n)} \tau^{(1+n)/2} 
\nonumber \\
& \times & \,  {_2}F_1 \left( 1, \frac{1+n}{2n} , \frac{1}{2} \left( 3+\frac{1}{n} \right) ; -\frac{a_{1,0}\tau^n}{a_{2,0}} \right)
\end{eqnarray}

Nevertheless, it is straightforward to express the scale factor as a function of the scalar field:
\begin{equation}
\label{powerscale}
a(\phi) = a_{1,0}e^{C\phi} + a_{2,0}e^{D\phi}
\end{equation}
where
\begin{equation}
\label{powerAB}
C=\sqrt{\frac{1+n}{2(1-n)}} , \qquad D= \sqrt{\frac{1-n}{2(1+n)}} , \qquad CD=\frac{1}{2}
\end{equation}
An expression for the scalar field potential can then be deduced by expressing the FRW equations in the `Hamilton-Jacobi' form, whereby the scalar field is interpreted as the effective dynamical variable \cite{GS1988,SBB1989,M1990,L1991,infreview}. (This is consistent if the field rolls monotonically with time and this constraint is manifestly satisfied for the class of models under consideration.) Specifically, when $k=0$, Eqs. (\ref{frw1})-(\ref{frw3}) may be written in the form 
\begin{eqnarray}
\label{HJ}
H'(\phi) &  = &-\frac{1}{2} \frac{d\phi}{dT} 
\\
a(\phi)H(\phi) & = & -2 a'(\phi)H'(\phi)
\\
\label{HJ3}
V(\phi) & = & 3H^2(\phi) -2 H'^2(\phi)
\end{eqnarray}
We then find that the Hubble parameter is given by 
\begin{equation}
\label{powerhubble}
H(\phi) = \hat{a}_1 e^{-C\phi} +\hat{a}_2 e^{-D\phi}
\end{equation}
where the constants satisfy the algebraic constraint
\begin{equation}
\label{powerconstraint}
(2C^2-1)a_{1,0}\hat{a}_1 +(2D^2-1)a_{2,0}\hat{a}_2 =0
\end{equation}
It then follows from Eq. (\ref{HJ3}) that the potential is given by a series of exponential terms together with a constant term.

\renewcommand{\thesection}{5}

\section{\label{virasorosection} Cosmology and the Virasoro Coadjoint Orbit}

Thus far, we have considered diffeomorphisms on the line. In this Section, we focus on the diffeomorphism group on the circle, $S^1$. More precisely, we consider its universal cover group $\diffc=\diff$ and the corresponding central extension, which we denote as $\diffext$. (See the Appendix for definitions and background details.) This will allow us to interpret a subset of scalar field cosmologies in terms of a coadjoint orbit of the Virasoro group. (For an introduction to the Virasoro group and its coadjoint representation, see \cite{O2006,KW2009,O2016,GR2007}.) 

Elements of $\diffc$ are $2\pi \mathbb{Z}$-equivariant diffeomorphisms $f \, : \, \mathbb{R} \rightarrow \mathbb{R}$ on the real line defined by $f (\varphi +2\pi)=f(\varphi)+2\pi$, $f'(\varphi) >0$ for any $\varphi \in \mathbb{R}$. Equivalently, we may write $f(\varphi) =\varphi +u(\varphi)$, where $u(\varphi) \in C^{\infty}(S^1)$ is a $2\pi$-periodic function on the circle. 
The Lie algebra is the tangent space at the identity corresponding to the space of infinitesimal diffeomorphisms, $f(\varphi)=\varphi + \epsilon X(\varphi)$, where $\epsilon$ is a small parameter and $X(\varphi) \in C^{\infty}(S^1)$. It is the space of smooth vector fields on $S^1$. We therefore regard $\diffc$ as a subgroup of ${\rm Diff}^+(\mathbb{R})$, where functions on the circle are $2\pi$-periodic functions on the real line $\mathbb{R}$. 

Before proceeding, it is worth remarking that in inflationary scenarios of the early universe in which the Lorentzian quasi-de Sitter expansion is postulated to emerge from an initial quantum state determined by a microcanonical density matrix, the gravitational instantons which dominate the path integral are defined as solutions to the semi-classical field equations on the positive-definite FRW metric $ds^2 = N^2(\tau)d\tau^2 +a^2(\tau) d\Omega^2_{3,k}$, where Eucliedean time $\tau$ is compactified on the circle, $S^1$ \cite{B2007,BZ2017}. Since the cosmological scalar field equations are form invariant under a Wick rotation to Euclidean time $t \mapsto it$ such that $V(\phi) \mapsto -V(\phi)$, $k \mapsto -k$, the present discussion is also relevant to these scenarios.

The coadjoint representation of the centrally extended Virasoro group is
defined by its action on the coadjoint vector $(p,c)$, where $p \in {\cal{F}}_2(S^1)$ is a quadratic density and $c \in \mathbb{R}$ is the central charge \cite{LP1975,K1980,S1981,K1982,W1988,B1988,BLP1998,GB2009}. It produces the image [see Eq. (\ref{adjoint}) in the Appendix]
\begin{equation}
\label{maincoad}
\left( \widehat{\rm Ad}^*_{\beta^{-1}} p \right) (\tau ) = ( \beta' (\tau) )^2 p (\beta(\tau)) -\frac{c}{12} S[\beta] (\tau)
\end{equation}
in terms of the inverse map $\varphi = \beta (\tau )= f^{-1} (\tau)$, where $S[\beta ](\tau)$ is the Schwarzian derivative (\ref{schwarzian}). It is important to note for the present discussion that the central charge remains invariant under the adjoint representation and, moreover, it introduces a non-trivial term into the transformation rule for the 
dual quadratic density, $p$. 

The adjoint map (\ref{maincoad}) generates an {\em identical} image (\ref{deftildelambda}) to that produced by the ${\rm Diff}({\cal{M}})$-actions (\ref{virfinite})-(\ref{virwave}) and (\ref{diffrep}) on the spaces of Schr\"odinger and Sturm-Liouville operators (\ref{defSlin}) and (\ref{SLoperator}), respectively (when we normalize to $m=-c/6$) \cite{K1980,S1981,K1982}. Hence, we may associate each coadjoint vector $(p,c)$ with a differential operator $\Delta_{\rm SL} = -\frac{c}{6} \partial^2_{\tau} +p(\tau)$, and vice-versa, where $\tau$ is the coordinate on the line and $p(\tau)$ is $2\pi$-periodic. The map associating each coadjoint vector with Hill's equation, $\Delta_{\rm SL} (X) =0$, is therefore $\diffc$-invariant. 

In this interpretation, we identify the scalar field's kinetic energy $K[\phi(\tau)] =(d\phi/d\tau)^2$ as the component 
of the quadratic density, $p$. Although this is periodic in $\tau$, it does not necessarily follow that the field's canonical kinetic energy expressed as a function of cosmic time is necessarily oscillatory. Indeed, solutions to Hill's equation need not be periodic and each model must be considered on a case-by-case basis. 

Following an identical argument to that presented in Section \ref{orbitsection}, 
we may now identify dynamical scalar field cosmologies with elements on the orbit of the coadjoint vector $(p,c)=(0,c)$. The orbit is bijective to the quotient manifold ${\cal{O}}_{(p)} \cong \diffc /{\rm Stab}_{(p,c)}$, where
the stabilizer is the kernel of the coadjoint representation of the Virasoro algebra. This is the derivative of (\ref{maincoad}) at the identity [cf. Eq. (\ref{stab})]:
\begin{equation}
\label{mainresulttiny}
\widehat{\rm ad}^*_X p(\varphi) = Xp'+2pX' -\frac{c}{12} X'''
\end{equation}
The kernel is spanned by vector fields satisfying $X'''=0$ when $p=0$. The only $2\pi$-periodic solution satisfying this equation is $X(\tau ) = {\rm constant}$, so the stabilizer group is the one-dimensional group of rigid translations.
The orbit is therefore \cite{W1988}
\begin{equation}
\label{cosmoorbit}
{\cal{O}}_{(0,c)} \cong \frac{\diff}{\mathbb{R}}\cong \frac{{\rm Diff}^+(S^1)}{S^1}
\end{equation}
In the classification of Hill operators by $SL(2,\mathbb{R})$ monodromy matrices, this corresponds to a non-degenerate parabolic orbit with zero winding number \cite{K1982,BLP1998,KW2009}.

This might seem counter-intuitive, since the image of the vector $(0,c)$ under the coadjoint representation is determined entirely by the Schwarzian operator, which is manifestly invariant under $SL(2, \mathbb{R})$. However, one must bear in mind that we have restricted to $2\pi \mathbb{Z}$-equivariant diffeomorphisms in this Section, so the potential term, $p(\tau )$, in Hill's equation must be $2\pi$-periodic. 

Further insight may be gained by means of the Kirillov-Kostant-Souriau (KKS) theorem \cite{K1975,K1970,S1970}. This states that every coadjoint orbit ${\cal{O}}_{(p)}$ of a Lie group $G$ is a (homogeneous) symplectic manifold associated with a non-degenerate, $G$-invariant, closed two-form:
\begin{equation}
\label{kks}
\omega_p ({\rm ad}^*_X p,{\rm ad}^*_Y p ) := \langle p, [X,Y] \rangle
\end{equation}
Non-degeneracy is defined by the condition that the kernel of the two-form, ${\rm ker} (\omega) := \{ v\in {T\cal{O}}_{(p)} | i_v \omega =0 \}$, be zero everywhere on the manifold. Here $i_v$ is the contraction with respect to $v$, $X$ and $Y$ are elements of the adjoint representation on the tangent space $T{\cal{O}}_{(p)}$ to the orbit and $\omega_p (\cdot,\cdot)$ is a real-valued function on the orbit, with a value given by Eq. (\ref{kks}) at each point $\tilde{p} \in {\cal{O}}_{(p)}$ \cite{W1988}. 

We now proceed to calculate the symplectic two-form on the orbit ${\cal{O}}_{(0,c)}$, following Stanford \& Witten \cite{W1988,SW2017}. The value of the two-form at a given image point (\ref{maincoad}) of the coadjoint vector $(0,c)$ is determined from the Lie bracket of the centrally extended algebra $\vectext$. This is given by Eqs. (\ref{vectextbracket})-(\ref{gf}). The tangent space to ${\cal{O}}_{(0,c)}$ is parametrized in terms of infinitesimal changes in $\beta$ such that $X(\tau)=d\beta^{-1}(\tau) = -d\beta(\tau)/\beta'(\tau)$ \cite{W1988,SW2017}. Eqs. (\ref{vectextbracket})-(\ref{adjoint}) then lead to an expression for the symplectic two-form given by 
\begin{eqnarray}
\label{kks1}
\omega =  - \frac{c}{48\pi} \int_0^{2\pi} d\tau \, \left[ \left( \frac{d\beta(\tau)}{\beta'(\tau)} \right)' \wedge \left( \frac{d\beta (\tau)}{\beta'(\tau)} \right)''\right.\nonumber \\ \left. -  2 S[\beta](\tau) \left( \frac{d\beta(\tau)}{\beta'(\tau)} \right) \wedge \left( \frac{d\beta(\tau)}{\beta'(\tau)} \right)' \right]
\end{eqnarray}
In this expression, $d$ represents an abstract exterior derivative, which operates on $\beta$ but not on the coordinate $\tau$ \cite{SW2017}. It therefore commutes with $\partial_{\tau}$ and it follows that the expression for the two-form simplifies to 
\begin{equation}
\label{kks2}
\omega = -\frac{c}{48\pi} \int_0^{2\pi} d \tau \left[ \left( \frac{d\beta'(\tau)}{\beta'(\tau)} \right) \wedge \partial_{\tau} \left( \frac{d\beta'(\tau)}{\beta'(\tau)} \right) \right] 
\end{equation}
Since the two-form (\ref{kks2}) is expressed entirely in terms of the logarithmic derivative $d\beta'/\beta' = d \log \beta'$, it is manifestly closed, $d \omega =0$. 

An infinitesimal action of $\diffc$ is given by $\epsilon (\tau)$ corresponding to a vector field $V_\epsilon = \epsilon (\tau) \partial_\tau$ acting on $\beta (\tau)$ such that $\delta \beta (\tau) = \epsilon (\tau) \beta'(\tau)$. Cartan's formula implies that the Lie derivative of a form along a vector field $V_\epsilon$ is given in general by ${\cal{L}}_{V_\epsilon} \omega =i_{V_\epsilon} d\omega +d(i_{V_\epsilon} \omega )$. Hence, to verify that the two-form $\omega$ is invariant under $\diffc$, i.e., to show that ${\cal{L}}_{V_\epsilon} \omega  =0$, it is sufficient to prove the existence of a function $H_{\epsilon}$ such that $i_{V_{\epsilon}} \omega =dH_{\epsilon}$. We take the contraction of $\omega$ with $V_\epsilon$ by replacing one of the $d\beta$ with $\delta \beta  = \epsilon \beta'$. Noting that 
$\partial_{\tau} (d\beta'/\beta') = d(\beta''/\beta')$, we find that such a function is 
\begin{eqnarray}
\label{Hepsilon}
H_{\epsilon} & = & -\frac{c}{48\pi} \int_0^{2\pi}d\tau \, \left[ 2 \epsilon' \frac{\beta''}{\beta'}+ \epsilon \left( \frac{\beta''}{\beta'} \right)^2 \right]
\nonumber \\
& = & \frac{c}{24\pi} \int_0^{2\pi} d\tau \,  \epsilon (\tau ) S[\beta](\tau)   
\end{eqnarray}

However, the two-form has a zero-mode and is therefore degenerate. To see this, consider the Fourier expansion of a $\diffc$ element in the neighourhood of the identity, $\beta (\tau)=\tau + \sum_{n \in \mathbb{Z}} u_ne^{ in\tau} $. At the identity ($u_n=0$), $\omega =\frac{ci}{24} \sum_n n^3 du_n\wedge du_{-n}$. Hence, the kernel of $\omega$ is one-dimensional, since $\omega$ vanishes for the $n=0$ mode. This is also true at all points on the orbit since $\omega$ is invariant under $\diffc$. 

This mode may be interpreted in the following way \cite{SW2017}. If we regard the $\diffc$ action of $V_\epsilon$ as a left-action, we may consider a second, right-action generated by a vector field $W_n$ given by $\delta \beta =e^{in\beta}$, $n \in \mathbb{Z}$. This right-action commutes with the left-action. It follows that $i_{W_n} \omega =-\frac{c}{24\pi}\int d\tau n^2 e^{in\beta}d\beta'$. Hence, $i_{W_n} \omega =0$ for the zero-mode $n=0$, which generates a translation symmetry. This implies that $\omega$ is a pullback from a quotient space that is diffeomorphic to $\diffc/\mathbb{R}$. Hence, once this zero mode has been removed by taking the quotient, $\omega$ represents a closed, non-degenerate two-form. We see, therefore, that the orbit ${\cal{O}}_{(0,c)}$ is indeed the quotient of $\diffc$ with the one-dimensional manifold $\mathbb{R}$, rather than with $SL(2,\mathbb{R})$.

Since $\omega$ is now non-degenerate, we may interpret function (\ref{Hepsilon}) as the symplectic generator of the (left-) action of the group. Indeed, the case $\epsilon (\tau)=1$ generates the analogue of a time translation symmetry $\delta\beta = \beta'$. The Hamiltonian generator in this case is a pure integral of the Schwarzian derivative 
\begin{equation}
\label{I}
I = \frac{c}{24\pi} \int d\tau \,  S[\beta](\tau)
\end{equation}
Within a cosmological context, this is directly related to the kinetic energy of the scalar field $K(\phi) = (d\phi /d\tau)^2$ via Eq. (\ref{defphi}):
\begin{equation}
\label{Gc/3}
I = \frac{Gc}{3} \int d \tau \, \left( \frac{d\phi}{d\tau} \right)^2
\end{equation}
Readers familiar with the Bondi-van der Burg-Metzner-Sachs (BMS) symmetry group \cite{B1960,BBM1962,S1962a,S1962b} of three-dimensional asymptotically flat gravity at null infinity will note the prefactor of $Gc/3$. A central charge of $c=3/G$ for the coadjoint representation of the centrally extended BMS group in three dimensions is precisely the condition that allows the metric components to be identified directly as components of a dual stress tensor in a two-dimensional field theory \cite{BC2007,BGG2012,P2017}.


\renewcommand{\thesection}{6}

\section{\label{concsection} Conclusion and Discussion}

In this paper, we have presented a new interpretation of FRW cosmologies in terms of the diffeomorphism group of the real line. The time reparametrization generates an effective rolling scalar field, where the kinetic energy of the field is related to the Schwarzian derivative of the diffeomophism. Specifically, we have considered a non-linear Schr\"odinger equation possessing a kinematical $SL(2,\mathbb{R})$ Lie point symmetry and employed the moment method to determine the dynamics of integral quantities describing the width, momentum and energy of the wavefunction. The ${\rm Diff}({\cal{M}})$-action explicitly breaks such a symmetry, but in such a way that the deformed evolution equations for the wavefunction moments can be mapped directly onto the (traceless) Einstein field equations of spatially isotropic universes. The wavefunction moments admit natural physical interpretations in terms of the scale factor of the universe (width of the wavefunction), the cosmic expansion rate (wavefunction momentum) and the energy density of the scalar field (Hamiltonian of the wavefunction). Energy-momentum conservation of the gravitational matter source (scalar field) arises naturally within the moment equations of motion and its origin can be traced to the existence of the conserved (in time) quadratic Casimir of the underlying $SL(2,\mathbb{R}) \cong SO(2,1)$ symmetry algebra. In effect, the $SL(2,\mathbb{R})$ symmetry is explicitly broken by the action of ${\rm Diff}({\cal{M}})$, but in such a way that energy-momentum conservation in the gravitational setting is maintained. 

We have focused on the one-dimensional, quintic NLS equation, which describes the dynamics of dilute, ultra-cold, strongly repulsive Bose gases interacting in the Tonks-Girardeau regime \cite{KNSX2000}. Such systems have been observed experimentally \cite{Petal2004,KWW2004}. It should be emphasized, however, that the moment method generates a closed system of evolution equations for a $d$-dimensional NLS equation with conformal (scale-invariant) non-linear coupling of the form $|u|^{4/d} u$ \cite{PTM2005}. The $d=2$ cubic NLS equation is of particular importance in the study of Bose-Einstein condensates which are tightly bound in one direction (see, e.g., \cite{Ueda2010} for a review). It should be straightforward to generalize the results of the present paper to such a higher-dimensional setting. 

The underlying $SL(2,\mathbb{R})$ symmetry is key to ensuring that the moment equations form a non-autonomous, closed system of ODEs, which ultimately reduce to an Ermakov-Pinney equation. Indeed, the symmetry is manifest as the Lie point symmetry (\ref{freeEPisometry}) of the free EP equation. This is important when considering the ${\rm Diff}({\cal{M}})$ actions on the (affine) spaces of Schr\"odinger and Sturm-Liouville operators with time-dependent (harmonic) potentials. Both spaces are preserved by natural representations of ${\rm Diff}({\cal{M}})$ and the corresponding NLS and EP equations are form-invariant under such an action. Moreover, the action of the diffeomorphism group on these operator spaces is {\em equivalent}, in the sense that a given time-dependent potential is mapped to the same image  (\ref{deftildelambda}) in both cases. In particular, this implies that the same trapping potential is generated in terms of the Schwarzian derivative from the $\lambda (t)=0$ model.

Consequently, the reduction of the non-linear Schr\"odinger equation to the EP equation via the moment method does not affect the image of the trapping potential under ${\rm Diff}({\cal{M}})$. The EP moment equation then maps directly onto the non-trivial component of the traceless Einstein field equations and the Casimir constraint provides the first integral constraint which ensures conservation of energy-momentum. 

The Schwarzian operator is the unique set of derivatives that remains invariant under a homographic transformation of the group of fractional linear transformations. As a result, each  scalar field cosmology (as parametrized by the field's kinetic energy) can be identified with an element of ${\rm Diff}({\cal{M}})$ modulo an $SL(2,\mathbb{R})$ time-reparametrization. A given cosmology is therefore represented by a point on the orbit of the element $(\lambda , m)=(0,m)$, which is the quotient space ${\rm Diff}({\cal{M}})/SL(2,\mathbb{R})$. A crucial point to note is that the image of the element $(0,m) \mapsto (\frac{m}{2} S[\beta](\tau) , m )$ is non-trivial because the mass parameter $m$ is non-zero. 

We interpreted the element $(0,m)$ as the constant Hamiltonian wavefunction corresponding to the de Sitter universe. The ${\rm Diff}({\cal{M}})$-action generates a self-similar wavefunction from this background configuration, which embeds the cosmic dynamics within its lowest-order moments. Self-similar wavefunctions generated by infinitesimal diffeomorphisms may then be identified as slow-roll inflationary cosmologies. Indeed, slow-roll inflation may be associated in this way with the explicit breaking of (non-relativistic) $SL(2,\mathbb{R}) \cong SO(2,1)$ conformal symmetry by a natural action of ${\rm Diff}({\cal{M}})$.

Focusing the analysis on the central extension of the universal cover group $\diffc$ of $2\pi \mathbb{Z}$-equivariant diffeomorphisms on the line leads to further insights. A rolling scalar field arises due to the non-zero central charge and the Schwarzian derivative which determines the field's kinetic energy is then the Souriau one-cocycle associated with the Bott-Thurston two-cocycle defining the central extension \cite{B1977,K1982}. Specific cosmologies may be identified as elements on the Virasoro coadjoint orbit ${\cal{O}}_{(0,c)}$. This symplectic manifold admits a Hamiltonian generator $H_{\epsilon} \sim \int d\tau (d\phi/d\tau)^2$ in terms of the scalar field kinetic energy. Intriguingly, the constant of proportionality is $cG/3$, a factor which equals unity in the ${\rm BMS}_3/{\rm CFT}_2$ correspondence \cite{BC2007,BGG2012,P2017}. 
In general, the transformation of the quadratic density $\lambda$ under the coadjoint action (\ref{maincoad}) may be interpreted as that of a stress-tensor in a two-dimensional CFT. In this context, the integral $E[\tilde{\lambda} ] := \frac{1}{2\pi} \int d \tau \tilde{\lambda} (\tau)$ defines the associated energy of the CFT. With $c=3/G$, such an interpretation leads to an energy functional 
\begin{equation}
\label{enfunc}
E= \int d\tau \left( \frac{d\phi}{d\tau} \right)^2
\end{equation}
for each inflationary cosmology. It would be interesting to investigate if there are possible connections with the three-dimensional BMS group.

We conclude with a speculative remark. Stanford \& Witten \cite{SW2017} have recently calculated the $\epsilon (\tau)={\rm constant}$ Hamiltonian generator associated with the Virasoro coadjoint orbit of the vector $(-c/24,c)$. This orbit is the quotient $\diffext /SL(2,\mathbb{R})$. They identified this generator with the Schwarzian action of the Sachdev-Ye-Kitaev (SYK) model \cite{SY1993,K2015}, which is a quantum-mechanical model of large $N$ fermionic particles with emergent reparametrization invariance. This model has attracted considerable interest recently in light of its conjectured duality with $(1+1)$-dimensional gravity \cite{MS2016,MSY2016,MTV2017,MNW2017,GMSVV2017,CRYZ2017,S2017}. At the infrared fixed point, the reparametrization is spontaneously broken to $SL(2,\mathbb{R})$, leading to zero modes (Goldstone bosons) that parametrize the quotient space ${\rm Diff}({\cal{M}})/SL(2,\mathbb{R})$. This is similar to what happens when one introduces a cutoff in two-dimensional anti-de Sitter space arising as the low-energy limit of some generic UV theory \cite{MSY2016}. The cutout can be parametrized in terms of a function $t (\tau)$, where $\tau$ is the time coordinate of the UV theory. However, the cutout shape is invariant under an $SL(2,\mathbb{R})$ reparametrization of $t$, so the full set of diffeomorphisms is ${\rm Diff}({\cal{M}})/SL(2,\mathbb{R})$. When one accounts for explicit boundary terms involving the dilaton, the shape of the cutout curve (reparametrization) is determined by an effective one-dimensional boundary action of the Schwarzian type, $I \sim \int d\tau \phi_r(\tau)S[t](\tau)$, where $\phi_r$ is determined by the boundary value of the dilaton \cite{MSY2016}. In view of the roles played in the present work by ${\rm Diff}({\cal{M}})/SL(2,\mathbb{R})$ and the Schwarzian derivative arising in the symplectic generator (\ref{I}), one might wonder if inflationary scalar field cosmology could be (formally) associated with lower-dimensional gravity in a similar way.  

\subsection*{Acknowledgements}

\noindent

We thank T. Harko, M. Lake, P. Moniz, and C. Terrero for numerous helpful communications and especially thank T. Harko and M. Lake for organizing and providing hospitality at the Cluj-Napoca Astronomical Observatory. The manuscript was prepared using Overleaf Latex.


\renewcommand{\thesection}{A}
\section{\label{appendix} Appendix: The Virasoro Group and the Coadjoint Representation}

The universal cover of the unit circle, $S^1=\{ e^{i\varphi} \in \mathbb{C} \, | \, \varphi\in  [ 0,2\pi ] \}$, is the real line $\mathbb{R}$ with the projection $\mathsf{p}:\mathbb{R} \rightarrow S^1 \, :\varphi \rightarrow  e^{i\varphi}$. The kernel of $\mathsf{p}$ is the set of integer multiples of $2\pi$, so $S^1  \cong \mathbb{R}/2\pi \mathbb{Z}$, where the quotient is by the equivalence $\varphi \sim \varphi +2\pi$. A diffeomorphism of the circle is a smooth bijection $F: S^1 \rightarrow S^1$, where composition determines the group operation. ${\rm Diff}(S^1)$ is the infinite-dimensional group of such diffeomorphisms. The connected subgroup of ${\rm Diff}(S^1)$ that preserves orientations on the circle is denoted ${\rm Diff}^+(S^1)$.  

In terms of the $2\pi$-periodic coordinate $\varphi$, an element of ${\rm Diff}(S^1)$ is the map $F: e^{i\varphi} \rightarrow F(e^{i\varphi})$. Given such a map, there exists a diffeomorphism $f: \mathbb{R} \rightarrow \mathbb{R}$ of the real line such that 
\begin{equation}
\label{projection}
F(e^{i\varphi})=e^{if(\varphi)}, \qquad F \circ \mathsf{p} = \mathsf{p} \circ f 
\end{equation}
To respect the periodicity of $\varphi$ and the orientation of $S^1$, the map  $f: \mathbb{R} \rightarrow \mathbb{R}$ must be a (smooth) $2\pi \mathbb{Z}$-equivariant map defined by the properties  (in this appendix a prime denotes $\partial/\partial \varphi$)
\begin{equation}
\label{defequiv}
f'(\varphi ) >0, \qquad f (\varphi +2\pi) = f(\varphi) +2\pi
\end{equation}
for any $\varphi \in \mathbb{R}$ on the line. Such a map can always be expressed in the form 
$f (\varphi)=\varphi +u(\varphi)$, where $u(\varphi)$ is a smooth $2\pi$-periodic function, i.e., a smooth function on the circle, $u(\varphi)\in C^{\infty}(S^1)$. 
The group of $2\pi \mathbb{Z}$-equivariant maps on the real line $\mathbb{R}$ defined by (\ref{defequiv}) with the projection (\ref{projection}) form the universal cover group of 
${\rm Diff}^+ (S^1)$. We focus on this universal cover group and denote it by $\diff = \diffc$. 

The coadjoint action of $\diffc$ is defined on the (regularized) space that is dual to $\vectc$, the space of smooth vector fields on the circle. The dual is the space of quadratic densities, $\vectc^* \cong \mathcal{F}_2(S^1) = \{ p(\varphi)(d\varphi)^2\, | \, p(\varphi) \in C^{\infty}(S^1) \}$. It is a dual space in the sense that there exists a non-degenerate pairing (an isomorphism  $\mathcal{F}_2(S^1)^*\cong \mathcal{F}_{-1}(S^1)$) that is invariant under the action of $\diffc$. The pairing is given explicitly by means of the $L^2$--inner product:
\begin{equation}
\label{vectdual}
\langle p, X\rangle : = \frac{1}{2\pi} \int_0^{2\pi} d \varphi \, p(\varphi) X(\varphi)  
\end{equation} 
The coadjoint representation of $\diffc$ is then defined to be ${\rm Ad}^* \,: \,  \diffc \rightarrow {\rm Aut} (\vectc^*)$ such that 
\begin{equation}
\label{coadjointG}
\langle {\rm Ad}_f^* \, p , Y \rangle := \langle p, {\rm Ad}_{f^{-1}} Y \rangle
\end{equation}
In other words, it is the transformation law for a quadratic tensor-density of weight $h=2$ [cf. Eq. (\ref{alphatrans})]: 
\begin{equation}
\label{adrepquad}
{\rm Ad}_f^*(p)(\varphi) = [(f^{-1})'(\varphi)]^2 p(f^{-1} (\varphi))
\end{equation}

In general, the central extension of a Lie group is defined in terms of a two-cocycle and an associated Souriau one-cocycle \cite{S1970}, with the cyclic property defined to ensure associativity of the extended group operation. The {\em Virasoro} group is defined here as the central extension of the universal cover group of the diffeomorphism group on the circle: 
$\diffext := \diffc \times \mathbb{R}$ \cite{O2006,KW2009,O2016,GR2007}. For this centrally extended group, the two-cocycle is the Bott-Thurston cocycle \cite{B1977} and the Souriau cocycle is the Schwarzian derivative \cite{S1981,K1982}. 
The Virasoro algebra $\vectext$ is the Lie algebra of $\diffext$. It is a vector space, $\vectext \cong \vectc \oplus \mathbb{R}$, spanned by pairs $(X, \mu )$, where $X =X(\varphi) \partial_{\varphi}$ is a vector field on $S^1$ and $\mu \in \mathbb{R}$. The Lie bracket of $\vectext$ is defined by 
\begin{equation}
\label{vectextbracket}
[(X,\mu),(Y,\nu) ]_c: =  ([X,Y], \, \mathsf{c}(X,Y) ) 
\end{equation}
where the commutator on the right-hand side of (\ref{vectextbracket}) is the Lie bracket of $\vectc$ and 
\begin{equation}
\label{gf}
\mathsf{c}(X,Y) :=   - \frac{1}{24\pi} \int_0^{2\pi} d\varphi \, X (\varphi) Y'''(\varphi)
\end{equation}
defines the Gelfand-Fuchs two-cocycle \cite{GF1968}.

The (regularized) dual Lie algebra $\vectext^*$ is therefore spanned by pairs $(p,c)$, where $p$ is a quadratic tensor-density on $S^1$ and $c \in \mathbb{R}$ is known as the central charge. The pairs $(X,\mu)$ and $(p,c)$ are referred to as adjoint and coadjoint vectors, respectively. The pairing between them is given by 
\begin{equation}
\label{virasoropairing}
\langle (p,c) ,(X, \mu ) \rangle := \frac{1}{2\pi} \int_0^{2\pi} d \varphi \, p(\varphi) X(\varphi) +c \mu
\end{equation}
The coadjoint representation of the centrally extended Virasoro group is then defined by \cite{O2006,KW2009,O2016,GR2007}
\begin{equation}
\nonumber 
\left\langle \widehat{\rm Ad}^*_{(f,\nu)} (p,c) ,(X, \mu) \right\rangle 
:= 
\left\langle (p,c), \widehat{\rm Ad}_{(f^{-1},-\nu)} (X, \mu ) \right\rangle 
\end{equation}
\begin{equation}
\label{coadextended}
=  
\left\langle p, 
{\rm Ad}_{f^{-1}} X \right\rangle +c \mu -\frac{c}{12} \left\langle \mathsf{S}[f^{-1}],X \right\rangle
\end{equation}
where the Souriau cocycle $\ \mathsf{S} [f^{-1}]$ is given by the Schwarzian derivative \cite{S1981,K1982,S1970}. Removing the explicit dependence on the adjoint vector $(X,\mu)$ from (\ref{coadextended}), we deduce the equivalent expression 
\begin{equation}
\label{adjoint}
\widehat{\rm Ad}_f^* (p,c) = \left( {\rm Ad}^*_f p - \frac{c}{12} \mathsf{S} [f^{-1}] ,c \right)
\end{equation}
for the coadjoint representation.

\end{document}